%% file: article.tex
\documentclass[11pt,epsf]{article}
 \usepackage{amsmath}
 \usepackage{graphicx}
 \usepackage[merge,numbers,compress]{natbib}
 \usepackage{xcolor} 
 \usepackage{dcolumn}
 \usepackage{amssymb}
 \usepackage[colorlinks=true,citecolor=blue!50!black,linkcolor=black]{hyperref}
 \usepackage{braket}
 \usepackage{subcaption}

\input{macros}

\begin{document}

\title{\hfill ~\\[-30mm]
\vspace{13mm}   \textbf{Perturbative Scalar M{\o}ller Scattering Simulation on an Ion-Trap Quantum Computer}}

\date{}
\author{
Herschel A.\ Chawdhry$^{1\,}$\footnote{E-mail:  \texttt{hchawdhry@fsu.edu}} ~and Mathieu Pellen$^{2\,}$\footnote{E-mail:  \texttt{mathieu.pellen@physik.uni-freiburg.de}}
\\[9mm]
{\small\it $^1$ Department of Physics, Florida State University,} \\ 
{\small\it 77 Chieftan Way, Tallahassee FL 32306, USA} \\[3mm] %
{\small\it $^2$ Albert-Ludwigs-Universit\"at Freiburg, Physikalisches Institut,} \\ %
{\small\it Hermann-Herder-Stra\ss e 3, D-79104 Freiburg, Germany}\\[3mm]
        }
\maketitle

\begin{abstract}
\noindent
We present quantum-computing algorithms to coherently simulate kinematic degrees of freedom in perturbative quantum field theory.
We focus in particular on scalar M{\o}ller scattering, where two identical scalar particles scatter via the exchange of a virtual scalar particle.
We present quantum circuits to simulate this process at tree level in 3+1 space-time dimensions.
We furthermore implement our algorithms in highly-optimised quantum circuits and use them to demonstrate perturbative QFT scattering on a state-of-the-art 98-qubit ion-trap quantum computer.
We find qualitative agreement with theoretical cross-section expectations and we study the impact of noise from the quantum hardware.
The work opens the door to using quantum computers to advantageously simulate scattering in perturbative quantum field theory.

\end{abstract}
\thispagestyle{empty}
\vfill

\newpage

\tableofcontents

\section{Introduction}

A frontline application of emerging quantum computers is the simulation of other quantum systems~\cite{osti_3914478}.
Within high-energy physics, the quantum simulation of perturbative Quantum Field Theory (QFT), including scattering amplitudes and cross sections, is of special interest because perturbative QFT calculations are the workhorse of precision physics predictions at high-energy particle colliders like the Large Hadron Collider (LHC), and so improved perturbative QFT predictions could enhance the discovery potential of collider experiments.
Along this line of research, quantum simulation techniques have been explored in many areas, ranging from numerical Monte Carlo integration~\cite{Bravo-Prieto:2021ehz,Agliardi:2022ghn,Robbiati:2023mbk,deLejarza:2023qxk,Cruz-Martinez:2023vgs,deLejarza:2024scm,Williams:2025hza,Lee:2025lzz} and parton-shower simulations~\cite{Bauer:2019qxa,Bepari:2020xqi,Bepari:2021kwv,Deliyannis:2022uyh,Gustafson:2022dsq,Bauer:2023ujy} to parton-distribution functions~\cite{Perez-Salinas:2020nem,Li:2021kcs} and fragmentation functions~\cite{deLejarza:2025upd}.
Furthermore, potential applications of quantum computing to perturbative QFT scattering amplitudes~\cite{Chawdhry:2023jks,Chawdhry:2025iuz,Bashore:2025uwb,Haddad:2026ncx,Bashore:2026ubq} and aspects of loop calculations~\cite{Ramirez-Uribe:2021ubp,Clemente:2022nll,deLejarza:2024pgk,Ramirez-Uribe:2024wua,Ochoa-Oregon:2025opz,Draper:2025nku} have also received significant attention in the literature.

A quantum-coherent simulation of perturbative QFT would be of particular interest because it could offer new routes to calculations by employing unique quantum computing capabilities including the manipulation of superpositions of quantum states.
As a first step in this direction, the quantum simulation of colour degrees of freedom in perturbative Quantum Chromodynamics (QCD) was studied in Refs.~\cite{Chawdhry:2023jks,Chawdhry:2025iuz}. The work showed early hints that coherent quantum simulations could potentially obtain an advantage by exploiting physical symmetries in a way that goes beyond the capabilities of classical computers.
Moving towards full scattering processes requires a direct and coherent implementation of kinematic degrees of freedom, which is the focus of the present article.
Past work on kinematic degrees of freedom~\cite{Chawdhry:2025iuz,Bashore:2025uwb,Haddad:2026ncx,Bashore:2026ubq} largely relied on quantum gates that incorporated classical precomputations of the kinematics, which somewhat restricted the prospects of obtaining a scalable quantum advantage over classical computers.

\begin{figure}[]
\begin{center}
\begin{subfigure}{0.40\textwidth}
\includegraphics[width=\textwidth]{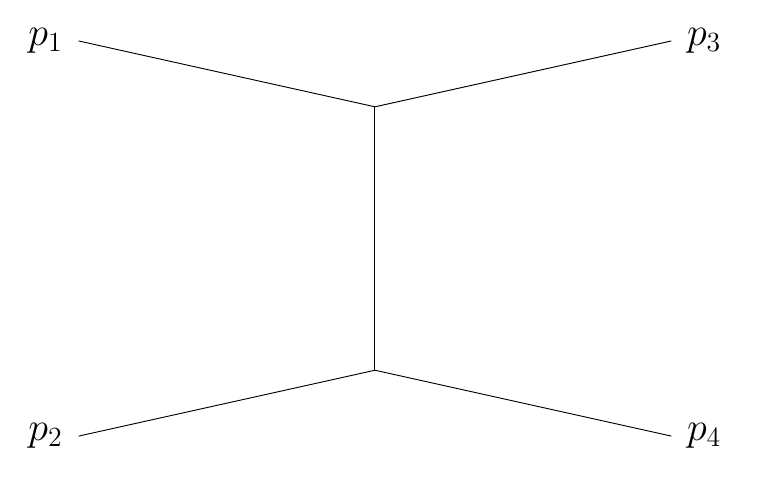}
\caption{\label{subfig:tchannel_diagram}$t$-channel Feynman diagram}
\end{subfigure}
\hspace{1cm}
\begin{subfigure}{0.40\textwidth}
\includegraphics[width=\textwidth]{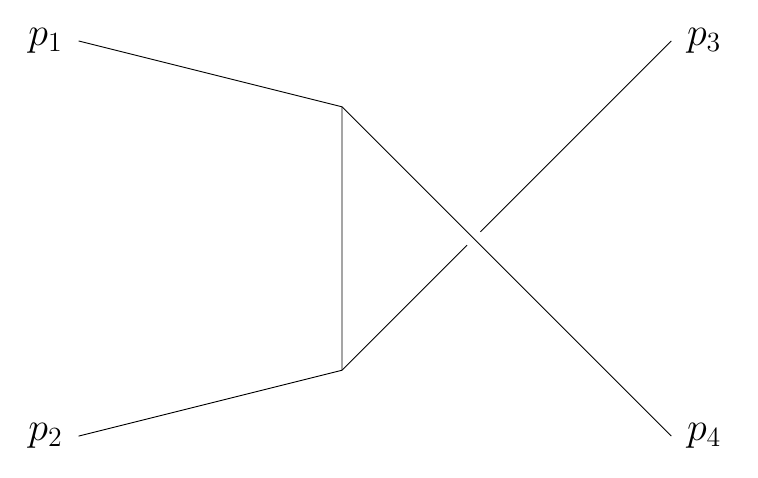}
\caption{\label{subfig:uchannel_diagram}$u$-channel Feynman diagram}
\end{subfigure}
\end{center}
\caption{\label{fig:diagrams_t_u_channels}
Tree-level Feynman diagrams for scalar M{\o}ller scattering.
}
\end{figure}

In this work, we present a direct and coherent quantum implementation of perturbative scalar field theory, and for concreteness we will consider scalar M{\o}ller scattering, specifically the elastic scattering of two massless scalar particles $\phi \phi \rightarrow \phi \phi$ via the exchange of another massless virtual scalar particle.
We have chosen this process because it allows us to demonstrate the essential components of our methods while keeping within the constraints of the best available present-day quantum computers.
We work at leading order in perturbative quantum field theory where the Feynman diagrams corresponding to this process are shown in Fig.~\ref{subfig:tchannel_diagram} and Fig.~\ref{subfig:uchannel_diagram}.
This scalar process can be seen as a prototype for M{\o}ller scattering $e^- e^- \rightarrow e^- e^-$ in Quantum Electrodynamics (QED) or $qq \rightarrow qq$ in QCD.
Note that our scattering demonstration is performed in the full 3 spatial dimensions.

Our methods coherently encode a quantum superposition of the Feynman diagrams into the wavefunction of the quantum computer such that the squared amplitude (including quantum interference effects between the Feynman diagrams) is obtained when the qubits are measured.
In performing this encoding we gain notable reductions in the quantum circuit depth by exploiting the permutation symmetry between the Feynman diagrams in a way that goes beyond the capabilities of classical computers.

We design compact quantum circuits to implement interaction vertices (including momentum conservation) and propagators.
The circuits employ an efficient discretisation of momenta which is well suited to perturbative QFT processes.
The size of the circuits furthermore scales logarithmically with the number of allowed momentum states, enabling rapid improvements in momentum precision as improved quantum computers become available in the future.
In the meantime, we have also highly optimised our circuits in order to make them amenable to present-day quantum computers.

We demonstrate our algorithms on a state-of-the-art 98-qubit ion-trap quantum computer with all-to-all connectivity between qubits.
We believe this is the first full demonstration of a perturbative QFT scattering process on a quantum computer.
Although our methods are scalable to arbitrary numbers of qubits, in our present demonstrations we employ only around 50 qubits in order to minimise the impact of hardware noise on the final phenomenological results.

The rest of this paper is organised as follows.
In Sec.~\ref{sec:momentum_and_phase_space} we present our momentum encoding and the phase-space generation.
In Sec.~\ref{sec:propagators} we present our implementation of propagators and use it to obtain a simulation of the $t$-channel Feynman diagram in Fig.~\ref{subfig:tchannel_diagram}.
In Sec.~\ref{sec:permuting_diagrams} we exploit the permutation symmetry between the two Feynman diagrams in order to generate a quantum superposition of them such that we obtain the fully-interfered squared amplitude (and hence cross-section) upon measuring the qubits encoding the scattering process.
In Sec.~\ref{sec:results} we demonstrate our algorithms on the 98-qubit ion-trap quantum computer.
Concluding remarks are given in Sec.~\ref{sec:conclusion}.

\section{Mometum encoding and phase-space generation}\label{sec:momentum_and_phase_space}

Since this work focusses on perturbative QFT rather than lattice QFT, we use qubits to encode the states of the \emph{field excitations} (i.e. particles) rather than directly encoding the fields themselves.
For each particle, we require a set of qubits.
In our previous work~\cite{Chawdhry:2023jks,Chawdhry:2025iuz} such qubits encoded the colour quantum numbers of the particle, whereas here the qubits encode the particle's momentum.
Our encoding is designed to allow a particle to be placed either into a state of definite momentum, or into a superposition of momentum states, which can furthermore be entangled with the states of other particles, just like in nature.

Working in $d=4$ space-time dimensions, we can represent the momentum state of a given particle by using $d$ registers, where each register comprises $n$ qubits.
Here $n$ is an integer controlling the precision to which the particle's momentum is represented: each $n$-qubit register can represent a (signed) $n$-bit integer in binary, or a quantum superposition of such integers.
Although this constitutes a discretisation of the momentum, note that it is very different from the discretisations employed in lattice calculations: the number of qubits required by a perturbative encoding to represent a given number of discrete momentum states is exponentially smaller than in a lattice-based encoding, albeit at the cost of a linear (rather than constant) scaling with the number of particles.
As will be seen in the rest of this section, this encoding allows for the implementation of propagators and interaction vertices (including momentum conservation) with quantum circuit sizes that scale efficiently (i.e. polynomially) with $n$.

Having defined our momentum encoding, our goal in this work will be to place the wavefunction of the quantum computer into an entangled superposition of momentum states that suitably represents the cross section $\sigma$ for the desired scattering process.
In general, for a process with 2 incoming particles with energies $E_1$ and $E_2$ and with velocities $\overrightarrow{v_1}$ and $\overrightarrow{v_2}$, the cross section is given by
\begin{equation}\label{eq:cross_section}
\sigma =
\frac{1}{4 E_1 E_2 |\overrightarrow{v_1} - \overrightarrow{v_2}|}
\int
|M|^2
d\Phi ,
\end{equation}
where $M$ is the Lorentz-invariant matrix element for the process and $\Phi$ is the Lorentz-invariant phase space of the outgoing particles.
For fixed initial conditions, the quantity $\frac{1}{4 E_1 E_2 |\overrightarrow{v_1} - \overrightarrow{v_2}|}$ is a constant and so it can be factored out of our quantum simulation.
The phase-space integral can be performed over the full range of $\Phi$ to obtain the total cross section, or it can be performed over phase-space bins to obtain differential cross-sections in one or more kinematic variables.

For concreteness, in the present work we will focus on the process of scalar M{\o}ller scattering, whereby two identical massless scalar particles $\phi$ scatter elastically via the exchange of another massless scalar particle:
\begin{equation}
\phi(p_1) + \phi(p_2) \rightarrow \phi(p_3) + \phi(p_4),
\end{equation}
where $p_i$ are the momenta of the particles.
We work for convenience in the following scalar field theory with a massless complex scalar $\phi$ interacting with a massless real scalar $\chi$:
\begin{equation}
\mathcal{L} = (\partial^\mu \phi)^{\dagger} (\partial_\mu \phi) + \frac{1}{2}(\partial^\mu \chi)(\partial_\mu \chi) + g  \chi \phi^{\dagger} \phi,
\end{equation}
where the coupling constant $g$ is assumed to be sufficiently small (relative to the momenta of the scattering process) so as to make the interaction perturbative.
At tree level, the scattering amplitude $M$ is given by the sum of the two Feynman diagrams shown in Figs.~\ref{subfig:tchannel_diagram} and~\ref{subfig:uchannel_diagram}.
Our scalar M{\o}ller scattering process can therefore be seen as a scalar version of the QED M{\o}ller scattering process $e^- e^- \rightarrow e^- e^-$.

Present-day quantum computers are limited in the number of qubits as well as by hardware noise, the latter imposing severe constraints on the number of consecutive quantum operations that can performed before quantum coherence is lost.
Therefore, in order to be able to demonstrate our example scattering process on real-world hardware, we set the momentum discretisation parameter to a very small value: $n=3$.
This allows each component of each momentum of each particle to take integer values in the range $\{-4, -3, -2, -1, 0, 1, 2, 3\}$.\footnote{An exception is the longitudinal component $k_z$ of the virtual exchanged particle, for which we had to set $n=4$ in order to allow a larger range of values $k_z \in \{-8, -7, \ldots, 6, 7\}$ required for our chosen kinematics.}
Note that, looking ahead to future quantum computers, this range of allowed momentum values would increase exponentially with $n$, while the total number of required qubits would grow only linearly with $n$, and the number of required quantum gates (as will be seen below) would grow only polynomially with $n$.

For the present scattering process, we set
\begin{equation}\label{eq:p1_and_p2}
p_1 = \begin{pmatrix} 3 \\ 0 \\ 0 \\ 3 \end{pmatrix}, \quad
p_2 = \begin{pmatrix} 3 \\ 0 \\ 0 \\ -3 \end{pmatrix}.
\end{equation}

For this choice of kinematics and discretisation, there are exactly 30 integer solutions satisfying momentum conservation
\begin{equation}
p_1^\mu + p_2^\mu = p_3^\mu + p_4^\mu
\end{equation}
and the on-shell conditions
\begin{equation}
p_i^2 = 0 \quad \forall i \in \{1,2,3,4\}.
\end{equation}
Of those 30 solutions, two correspond to forward scattering $\overrightarrow{p_3} = (0,0,\pm 3)$, where the tree-level matrix element diverges, so we define a cut
\begin{equation}
|p_{3,z}| < 2.5
\end{equation}
to exclude them.
In a realistic collider environment, such a cut would be replaced by a rapidity cut.

For phase-space generation, a generic quantum circuit is beyond the scope of the present work but we have hand-designed a compact quantum circuit for generating an equal superposition of the 28 allowed values of $\overrightarrow{p_3}$ compatible with the above kinematics:
\begin{equation}\label{eq:p3_superposition}
\frac{1}{\sqrt{28}}\sum_{\overrightarrow{p_3} \in \Phi_3} \ket{\overrightarrow{p_3}},
\end{equation}
where we write $\Phi_3$ to denote the set of 28 allowed vectors $\overrightarrow{p_3}$, and for any vector $\overrightarrow{q}$ we use the shorthand $\ket{\overrightarrow{q}}$ to denote a composite register comprising a separate register of qubits to store each component of $\overrightarrow{q}$:
\begin{equation}
\ket{\overrightarrow{q}} \equiv \ket{q_x}\ket{q_y}\ket{q_z}.
\end{equation}
In general the energy components of momentum should be represented in qubits just like the spatial components, but for our particular example all the energy components are trivially constant: $p_3^0 = p_4^0 = 3$ independent of the directions of the scattered particles.
The energy component $k^0$ of the virtual exchanged particle momentum $k$ in each of the 2 Feynman diagrams also takes a constant value: $k^0 = 0$.
Therefore, for our example process, we have reduced the number of required qubits and quantum gates by hardcoding the energy components of all particles as constant values in our circuits rather than representing them as variables in qubit registers.

Having created the state in Eq.~\eqref{eq:p3_superposition}, we can then use momentum conservation at interaction vertices to compute $k$ and $p_4$ via arithmetical quantum circuits in a cheap (i.e. polynomial in $n$) number of operations.
To begin, consider the upper vertex of the $t$-channel Feynman diagram. Since $p_1$ is known from eq.~\eqref{eq:p1_and_p2} and $p_3$ has already been generated by our phase-space circuit, we can build a quantum gate that performs the operation
\begin{equation}\label{eq:p3_to_p3kt}
\ket{p_3^\mu}\ket{0} \rightarrow \ket{p_3^\mu}\ket{k_{t-\textrm{channel}}^\mu},
\end{equation}
where
\begin{equation}
k_{t-\textrm{channel}}^\mu = p_3^\mu - p_1^\mu.
\end{equation}
The gate must be performed for each value of $\mu$, so as to calculate all components of $k_{t-\textrm{channel}}^\mu$,
although note that since the $\mu=0$ component is trivial as discussed above, we have optimised it out of our hardware implementation.
Since the gate in Eq.~\eqref{eq:p3_to_p3kt} is a quantum gate, applying it (for each value of $\mu$) to the 28-state superposition of $\ket{p_3}$ states in Eq.~\eqref{eq:p3_superposition} yields an entangled state
\begin{equation}\label{eq:superposition_p3_k}
\frac{1}{\sqrt{28}}\sum_{\overrightarrow{p_3} \in \Phi_3} \ket{\overrightarrow{p_3}}\ket{\overrightarrow{p_3} - \overrightarrow{p_1}},
\end{equation}
where the first register stores $\overrightarrow{p_3}$, the second register stores $\overrightarrow{k}_{t-\textrm{channel}}$.

Next, we use a similar procedure to implement the lower vertex of the $t$-channel Feynman diagram and thus write $\overrightarrow{p_4}$ into a new register: since
\begin{equation}
p_4^\mu = p_2^\mu - k_{t-\textrm{channel}}^\mu,
\end{equation}
we build a gate 
\begin{equation}\label{eq:k_to_kp4}
\ket{k^\mu_{t-\textrm{channel}}}\ket{0} \rightarrow \ket{k_{t-\textrm{channel}}^\mu}\ket{p_4^\mu}.
\end{equation}

Applying this operation to the state produced in Eq.~\eqref{eq:superposition_p3_k}, we obtain
\begin{equation}\label{eq:superposition_p3_k_p4}
\frac{1}{\sqrt{28}}\sum_{\overrightarrow{p_3} \in \Phi_3} \ket{\overrightarrow{p_3}}\ket{\overrightarrow{k}_{t-{\textrm{channel}}}}\ket{\overrightarrow{p_4}}
\end{equation}

To generalise beyond our present example, let us mention that for any tree-level amplitude, once the phase-space of the external particles has been generated, the momenta of intermediate virtual particles in a Feynman diagram can always be calculated cheaply (i.e. polynomially in $n$) using momentum-conserving interaction vertex gates like the one in Eq.~\eqref{eq:p3_to_p3kt}.
In general, this means encoding the momenta of all external and internal particles into qubits.
For our present example, however, representing $p_1$ and $p_2$ in qubits was unnecessary because their values are constant as defined in Eq.~\eqref{eq:p1_and_p2}.
In our above construction we therefore saved on qubits (and operations) by hardcoding the values of $p_1$ and $p_2$ into the relevant quantum gates, albeit at the cost of having to design separate quantum circuits for each vertex: Eq.~\eqref{eq:p3_to_p3kt} and Eq.~\eqref{eq:k_to_kp4}.
In the more general case, Eq.~\eqref{eq:p3_to_p3kt} can be generalised to a generic 3-point interaction vertex gate
\begin{equation}\label{eq:p1p3_to_p1p3kt}
\ket{p_1^\mu}\ket{p_3^\mu}\ket{0} \rightarrow \ket{p_1^\mu}\ket{p_3^\mu}\ket{k_{t-\textrm{channel}}^\mu}
\end{equation}
where $p_1$ is no longer hard-coded, and so the same circuit can be applied to a 3-point vertex in an arbitrarily complicated tree-level Feynman diagram.

The state in Eq.~\eqref{eq:superposition_p3_k_p4} is interesting because in the absence of any further operations, a hypothetical measurement of the qubits at this point would collapse the wavefunction and produce a pseudo-scattering event, for which the external momenta $\overrightarrow{p_3}$ and $\overrightarrow{p_4}$ (as well as the unphysical momentum $\overrightarrow{k}_{t-{\textrm{channel}}}$) could be read off in binary from the qubits that encode them.
For the state in Eq.~\eqref{eq:superposition_p3_k_p4}, there are 28 possible pseudo-scattering events and they would each be produced with equal probability $\frac{1}{28}$.
However, we want to obtain genuine scattering events, whose probability distribution matches the cross-section as defined in Eq.~\eqref{eq:cross_section}, and so we now need to implement the propagators (see Sec.~\ref{sec:propagators}) and the coherent summation of both Feynman diagrams (see Sec.~\ref{sec:permuting_diagrams}) including quantum interference effects.

\section{Block-encoding of propagators}\label{sec:propagators}

A generic scalar Feynman propagator $\frac{1}{q^2-m^2}$ for an internal particle with 4-momentum $q$ and mass $m$ can be implemented as follows.
We assume the momentum $q$ has already been encoded into a register, as is the case for $k_{t-\textrm{channel}}$ in the state in Eq.~\eqref{eq:superposition_p3_k_p4}.
We start by introducing a new register and then build a quantum circuit to encode $q^2-m^2$ into this register.
This can be done arithmetically in a time that is polynomial in the momentum-discretisation parameter $n$.
For our chosen example model we have mass $m=0$ but the inclusion of a non-zero mass if desired would add no additional complications.
For our chosen process and kinematics, $k^2_{t-\textrm{channel}}$ can take values from 6 to 30 and so we use a 5-qubit register which can store 5-bit unsigned integer values from 0 to 31.
It is straight-forward to build a gate to arithmetically perform the operation
\begin{equation}
\ket{\overrightarrow{q}}\ket{0} \rightarrow \ket{\overrightarrow{q}}\ket{q^2}.
\end{equation}
For our demonstration on a physical ion-trap quantum computer, we have performed additional optimisations exploiting the structure of the phase-space to reduce the number of quantum gates and hence the amount of hardware noise in our results.

Having encoded $q^2$ (or more generally, $q^2-m^2$) into a register of qubits, we now proceed to present an efficient block-encoding of the operation
\begin{equation}\label{eq:generic_propagator}
\ket{D} \rightarrow \frac{1}{D} \ket{D},
\end{equation}
where for the scalar case we have $D = q^2 - m^2$.
This will allow us to encode the \emph{scattering amplitude} of the QFT process into the \emph{probability amplitude} of a given configuration of qubits.

Our implementation of Eq.~\eqref{eq:generic_propagator} builds on the standard identity
\begin{equation}\label{eq:propagator_identity}
\frac{1}{X+i\epsilon} = -i \int_{0}^{\infty}d\lambda e^{i \lambda (X+i\epsilon)},
\end{equation}
where $X \in \mathbb{R}$.
Since phase rotation gates are readily available on present-day quantum computers, we can block-encode Eq.~\eqref{eq:generic_propagator} via phase rotations applied to the state $\ket{D}$.
Our discretisation of momenta and hence of $q^2$ allows for the phase rotations to be performed with a number of operations that is linear in $n$ and therefore scales very efficiently with increasing momentum precision.

Note that although we have chosen to work with Feynman propagators, it is straight-forward to adapt our implementation to apply instead to the advanced or retarded propagators $\frac{1}{E-H_0 \pm i\epsilon}$ of time-dependent perturbation theory. In that case, the $\lambda$ parameter takes the role of time, and the phase rotations $e^{i (E-H_0 \pm i\epsilon) \lambda}$ applied to the qubits will directly implement the time-evolution of the particle states.

We start by introducing a register $L$ to encode $\lambda$ in binary as an integer in the range $\{0, \ldots, \Lambda-1\}$ using $\log_2(\Lambda)$ qubits, where $\Lambda$ is a power of 2. For our hardware demonstrations we choose $\Lambda=32$, so the $L$ register has 5 bits, but a larger value for $\Lambda$ can be chosen in the future.
Initially, the state of the registers encoding $D$ and $\lambda$ is:
\begin{equation}
\ket{D}\ket{0}_L.
\end{equation}
We now apply a Hadamard gate to each qubit in the $L$ register and hence obtain the state
\begin{equation}
\frac{1}{\sqrt{\Lambda}}\sum_{\lambda=0}^{\Lambda-1} \ket{D}\ket{\lambda}.
\end{equation}
Next we introduce a continuum parameter $\theta$ (choosing  $\theta = 0.01$ for our hardware demonstrations) and apply phase rotations to obtain
\begin{equation}
\frac{1}{\sqrt{\Lambda}}\sum_{\lambda=0}^{\Lambda-1} e^{i D \lambda \theta} \ket{D}\ket{\lambda}.
\end{equation}
Crucially, this phase rotation is very efficient: we construct it from controlled-phase rotations applied to all pairings of qubits in the two registers.
We now introduce a constant $\epsilon$ satisfying
\begin{equation}\label{eq:epsilon_constraint}
\epsilon \ll D \quad \forall D.
\end{equation}
For our present kinematics, we have $D \geq 6$ at all phase-space points and so we choose to set $\epsilon = 3$.
Since multiplying a state by $e^{-\epsilon \lambda}$ is a non-unitary operation, we introduce an ancilla register $\mathcal{U}$ of the same size as the $L$ register and then apply $\log_2(\Lambda)$ rotation operations $\texttt{CY}$ to corresponding pairs of qubits in the $L$ and $\mathcal{U}$ registers to obtain:
\begin{equation}
\frac{1}{\sqrt{\Lambda}}\sum_{\lambda=0}^{\Lambda-1} e^{i D \lambda \theta} e^{- \epsilon \lambda \theta} \ket{D}\ket{\lambda} \ket{0}_\mathcal{U} + (\textrm{terms } \perp \textrm{ to } \ket{0}_\mathcal{U}).
\end{equation}
This state is mathematically equivalent to
\begin{equation}
\frac{1}{\sqrt{\Lambda}}\sum_{\lambda=0}^{\Lambda-1} e^{i (D + i \epsilon) \lambda \theta} \ket{D}\ket{\lambda} \ket{0}_\mathcal{U} + (\textrm{terms } \perp \textrm{ to } \ket{0}_\mathcal{U}).
\end{equation}
Finally, applying a Hadamard gate to each qubit in the $L$ register, we obtain the state
\begin{equation}
\frac{1}{\Lambda}\left(
\sum_{\lambda=0}^{\Lambda-1} e^{i (D + i \epsilon) \lambda \theta}
\right)
\ket{D}\ket{0}_L \ket{0}_\mathcal{U}
+ (\textrm{terms } \perp \textrm{ to } \ket{0}_L \ket{0}_\mathcal{U}).
\end{equation}
The previous equation contains a geometric series:
\begin{align}
\sum_{\lambda=0}^{\Lambda-1} e^{i (D + i \epsilon) \lambda \theta}
&=
\frac{e^{\Lambda \theta i (D+i\epsilon)}-1}{e^{ \theta i (D+i\epsilon)}-1}
\\ &\approx  \frac{i}{\theta} \frac{1}{D+i\epsilon}
\end{align}
where the final line assumes Eq.~\eqref{eq:epsilon_constraint} as well as:
\begin{equation}
\Lambda \theta \epsilon \gg 0 \quad \textrm{and} \quad \theta \ll \frac{1}{D_{\textrm{min}}}.
\end{equation}
We thus obtain a discretised block-encoding of the identity~\eqref{eq:propagator_identity} as desired.

Applying this block-encoding method to the state in Eq.~\eqref{eq:superposition_p3_k_p4}, we obtain a full encoding of the $t$-channel amplitude
\begin{equation}\label{eq:wavefunction_tchannel}
\mathcal{N} \sum_{\overrightarrow{p_3} \in \Phi_3}
\frac{1}{q^2}
\ket{\overrightarrow{p_3}}
\ket{\overrightarrow{p_4}}
\ket{\overrightarrow{q}}
\ket{q^2}
\ket{0}_L
\ket{0}_\mathcal{U}
+ (\textrm{terms } \perp \textrm{ to } \ket{0}_L \ket{0}_\mathcal{U}),
\end{equation}
where $q$ = $k_{t-{\textrm{channel}}}$ and we have a normalisation $\mathcal{N} = \frac{i}{\Lambda \theta}$.

\section{Generating permuted Feynman diagrams and interferences}\label{sec:permuting_diagrams}

Having obtained an encoding of the $t$-channel amplitude in Eq.~\eqref{eq:wavefunction_tchannel}, we can uncompute the registers containing $\ket{\overrightarrow{q}}$ and $\ket{q^2}$ to get
\begin{equation}\label{eq:wavefunction_tchannel}
\mathcal{N} \sum_{\overrightarrow{p_3} \in \Phi_3}
\frac{1}{q^2}
\ket{\overrightarrow{p_3}}
\ket{\overrightarrow{p_4}}
\ket{0}
\ket{0}
\ket{0}_L
\ket{0}_\mathcal{U}
+ (\textrm{terms } \perp \textrm{ to } \ket{0}_L \ket{0}_\mathcal{U}).
\end{equation}
This encoded form of the Feynman diagram is very convenient because it allows us to apply the swaps-based technology of Ref.~\cite{Chawdhry:2025iuz} to generate permuted Feynman diagrams.

As a preliminary, let us note that if we wanted the $u$-channel diagram alone, we could obtain it from the $t$-channel diagram by performing a swap operation on the qubits encoding $\ket{\overrightarrow{p_3}}$ and $\ket{\overrightarrow{p_4}}$:
\begin{equation}\label{eq:swap_p3_p4}
\ket{\overrightarrow{p_3}}
\ket{\overrightarrow{p_4}}
\rightarrow
\ket{\overrightarrow{p_4}}
\ket{\overrightarrow{p_3}}
\end{equation}
Since for our chosen example, $\ket{\overrightarrow{p_3}}$ and $\ket{\overrightarrow{p_4}}$ each contain $9$ qubits, the swap~\eqref{eq:swap_p3_p4} would require 9 bitwise swaps.

A far more useful result, however, is if we obtain a superposition of the $t$- and $u$-channel diagrams. We introduce an ancilla qubit $c$ and use a Hadamard gate to place it into a state $\frac{\ket{0}+\ket{1}}{\sqrt{2}}$.
If we now perform the swap in Eq.~\eqref{eq:swap_p3_p4} but control it on the qubit $c$ being in the state $\ket{1}$, we obtain the state
\begin{equation}\label{eq:wavefunction_cswapped}
\mathcal{N}
\sum_{\overrightarrow{p_3} \in \Phi_3}
\frac{1}{k^2_{t-{\textrm{channel}}}}
\left(
\frac{
\ket{\overrightarrow{p_3}}
\ket{\overrightarrow{p_4}}
\ket{0}_c
+
\ket{\overrightarrow{p_4}}
\ket{\overrightarrow{p_3}}
\ket{1}_c
}{\sqrt{2}}
\right)
\ket{0}
\ket{0}
\ket{0}_L
\ket{0}_\mathcal{U}
+ (\textrm{terms } \perp \textrm{ to } \ket{0}_L \ket{0}_\mathcal{U}).
\end{equation}

Now applying another Hadamard gate to the qubit $c$, we obtain
\begin{equation}\label{eq:wavefunction_hadamardc}
\mathcal{N}
\sum_{\overrightarrow{p_3} \in \Phi_3}
\frac{1}{k^2_{t-{\textrm{channel}}}}
\left(
\frac{
\ket{\overrightarrow{p_3}}
\ket{\overrightarrow{p_4}}
+
\ket{\overrightarrow{p_4}}
\ket{\overrightarrow{p_3}}
}{2}
\right)
\ket{0}_c
\ket{0}
\ket{0}
\ket{0}_L
\ket{0}_\mathcal{U}
+ (\textrm{terms } \perp \textrm{ to } \ket{0}_c\ket{0}_L \ket{0}_\mathcal{U}).
\end{equation}
This is mathematically equivalent to
\begin{equation}\label{eq:wavefunction_t_u}
\mathcal{N}
\sum_{\overrightarrow{p_3} \in \Phi_3}
\frac{1}{2}
\left(
\frac{1}{k_{t\textrm{-channel}}^2} + \frac{1}{k_{u\textrm{-channel}}^2}
\right)
\ket{\overrightarrow{p_3}}
\ket{\overrightarrow{p_4}}
\ket{0}_c
\ket{0}
\ket{0}
\ket{0}_L
\ket{0}_\mathcal{U}
+ (\textrm{terms } \perp \textrm{ to } \ket{0}_c\ket{0}_L \ket{0}_\mathcal{U}),
\end{equation}
where $k^\mu_{u-{\textrm{channel}}} = p_4^\mu-p_1^\mu$ is the momentum flowing through the exchanged particle in the $u$-channel diagram Fig.~\ref{subfig:uchannel_diagram}.

Therefore, if we measure the state of the quantum computer and select on the state $\ket{0}_c \ket{0}_L \ket{0}_\mathcal{U}$, we will obtain \emph{unweighted events} in which we can directly read out the binary-encoded momenta ($\overrightarrow{p_3}, \overrightarrow{p_4}$).
These events will follow a statistical distribution
\begin{equation}
P\left(
\ket{\overrightarrow{p_3}},
\ket{\overrightarrow{p_4}}
\right)
\propto
\left|
\frac{1}{k_{t\textrm{-channel}}^2} + \frac{1}{k_{u\textrm{-channel}}^2}
\right|^2
\end{equation}
that directly reflects the phase-space $\Phi_3$ and squared amplitude $|M|^2$, including all quantum interference effects between the $t$-channel diagram and the $u$-channel diagram.

\section{Results on ion-trap quantum computer}\label{sec:results}

To demonstrate our algorithms and chosen example on a real-world quantum computer, we implemented our circuits in the Guppy language and ran them on the 98-qubit ion-trap Quantinuum Helios-1 machine, which is the largest available quantum computer that natively supports operations between arbitrary pairs of qubits.

We ran 14822 shots of our circuit. At the end of each shot we measure all of the registers: $(\overrightarrow{p_3},\overrightarrow{p_4},\overrightarrow{k},k^2,L,\mathcal{U},c)$.
At this stage we perform some basic consistency checks to validate the shot and diagnose obvious instances of hardware noise.
In particular, we expect $\overrightarrow{k}=\overrightarrow{0}$ and $k^2 = 0$ because the circuit includes an uncomputation of those registers, and we also expect $\overrightarrow{p_1}+\overrightarrow{p_2} = \overrightarrow{p_3} + \overrightarrow{p_4}$ since all of our circuit components (phase-space generation, interaction vertices, propagators) are designed not to violate momentum conservation.
In practice, since we are running our circuits directly on physical qubits, we observe significant hardware noise: in nearly 50\% of the shots we see a violation of at least one of these consistency checks and so we discard the shot.
As will be seen in the results shortly to be discussed, this level of hardware noise can introduce significant biases and systematic uncertainties that are challenging to predict \emph{a priori}, so in the future we expect to derive significant benefit from improved hardware, further optimisation of circuits, and the implementation of quantum error correction and mitigation codes.

\begin{figure}[]
\center
\includegraphics[width=0.6\textwidth]{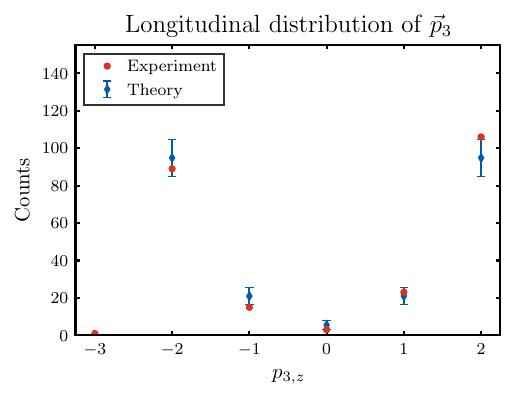}
\caption{\label{fig:longitudinal_distribution}%
Longitudinal momentum distribution of one of the outgoing particles.
The Feynman diagram in Fig.~\ref{subfig:tchannel_diagram} is directly simulated, while the Feynman diagram in Fig.~\ref{subfig:uchannel_diagram} is obtained by performing controlled swap operations on the qubits encoding the momenta of the two outgoing particles.
The output of the circuit provides the squared sum of the two Feynman diagrams, including interference effects.
Experimental results have been filtered to discard events containing obvious bit-flip errors, leading to only 50\% of events being displayed.
Theoretical predictions have been rescaled by a factor of 0.5 to enable comparison with experimental results.
Statistical uncertainties have been estimated for the rescaled theoretical predictions.
}
\end{figure}

\begin{figure}[]
\center
\includegraphics[width=0.6\textwidth]{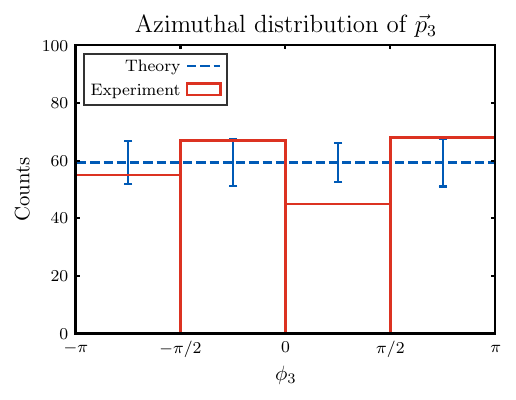}
\caption{\label{fig:azimuthal_distribution}%
Same as in Fig.~\ref{fig:longitudinal_distribution} but for the azimuthal-angle (i.e. $\phi_3$) distribution, where $\phi_3$ is defined such that $p_{3,x} = p_{3,T} \cos(\phi_3)$ and $p_{3,y} = p_{3,T} \sin(\phi_3)$, where $p_{3,T}$ is the transverse momentum of the particle. Events have been binned into four quadrants.
}
\end{figure}

After applying these consistency checks as well as post-selection on the $L$, $\mathcal{U}$, and $c$ registers, we obtain 237 events.
For each event, we directly read out the 9 qubits encoding $\overrightarrow{p_3}$ and the 9 qubits encoding $\overrightarrow{p_4}$, using the binary encoding explained in Sec.~\ref{sec:momentum_and_phase_space}.
The longitudinal (i.e. $p_{3,z}$) distribution of the events is shown in Fig.~\ref{fig:longitudinal_distribution} and displays good qualitative agreement with theoretical expectations.
The latter are calculated at the discretised phase space-points described in Sec.~\ref{sec:momentum_and_phase_space}.
Some possible noise effects in the experimental results are visible including indications of an asymmetry between the forwards and backwards events.
The azimuthal distribution of the events, binned into quadrants, is shown in Fig.~\ref{fig:azimuthal_distribution} and displays some indications of an anisotropy, which we ascribe to hardware noise effects.
Surprisingly, on one occasion a backwards event $\overrightarrow{p_3}=(0,0,-3)$, $\overrightarrow{p_4}=(0,0,+3)$ was observed even though our circuits are designed to only generate events with $|p_{3,z}| < 2.5$; we ascribe the presence of this event to hardware noise but for transparency we include it in Fig.~\ref{fig:longitudinal_distribution} because it passes the consistency checks that we defined in advance above.

Overall, despite the presence of hardware noise, we find that our results display good qualitative agreement with analytic expectations, which provides a validation of our methods, lays the groundwork for further simulations of continuous degrees of freedom in perturbative QFT, and motivates further algorithm development as quantum-computing technology continues to advance.

\section{Conclusion}\label{sec:conclusion}

In conclusion, we have presented and demonstrated quantum algorithms for the direct, coherent simulation of kinematic degrees of freedom in perturbative quantum field theory.
Our algorithms and qubit encodings of particle momenta allow for the direct implementation of propagators in Feynman diagrams, and they also enable the exploitation of physical symmetries in a way that goes beyond the capabilities of classical computers.

We used a state-of-the-art ion-trap quantum computer to perform a demonstration of our algorithms, using the example of scalar M{\o}ller scattering whereby two scalar particles scatter via the exchange of another scalar particle.
Our results for the kinematic distributions of the particles show good qualitative agreement with expectations, despite some hardware noise effects, and can serve as groundwork for future quantum algorithms for perturbative QFT calculations. With a modest increase in qubit counts, the latter could incorporate the kinematics of particles with spin and colour degrees of freedom as well as their interactions.

Our work also opens unique possibilities to use perturbative QFT-based techniques to verify and validate quantum computing hardware performance and benchmark its progress, including via physical symmetries, conservation laws, quantum interference between Feynman diagrams, and entanglement between particles.

Looking ahead, we expect that our coherent encoding of kinematics can provide a stepping stone for the development of further algorithms exploiting the qualitatively-new information processing capabilities of quantum computers in order to better understand and predict the phenomenology of perturbative QFT processes.

\section*{Acknowledgements}

We thank Christian Bauer, Fernando Febres Cordero, Takemichi Okui, and Laura Reina for helpful discussions.
The research of HAC is supported by the U.S. Department of Energy under grant DE-SC0010102.
This work was performed in part at the Aspen Center for Physics, which is supported by a grant from the Simons Foundation (1161654, Troyer).
This research was supported in part by grant NSF PHY-2309135 to the Kavli Institute for Theoretical Physics (KITP).
This research was supported by the Munich Institute for Astro-, Particle and BioPhysics (MIAPbP), which is funded by the DFG under Germany’s Excellence Strategy - EXC-2094 - 390783311.
An award of computer time was provided by the QUEST program. This research used resources of the Oak Ridge Leadership Computing Facility at the Oak Ridge National Laboratory, which is supported by the Advanced Scientific Computing Research programs in the Office of Science of the U.S. Department of Energy under Contract No. DE-AC05-00OR22725.


\bibliographystyle{utphys.bst}

\bibliography{biblio}
\end{document}

%% file: macros.tex
\def\be{\begin{equation}}
\def\ee{\end{equation}}

\newcommand{\newc}{\newcommand}
\newc{\bi}{\begin{itemize}}
\newc{\ei}{\end{itemize}}
\newc{\benu}{\begin{enumerate}}
\newc{\eenu}{\end{enumerate}}
\newc{\bc}{\begin{center}}
\newc{\ec}{\end{center}}
\newc{\bfig}{\begin{figure}}
\newc{\efig}{\end{figure}}
\newc{\qbar}{\bar{q}}
\newc{\go}{\tilde{g}}
\newc{\PB}{\textsc{Powheg-Box}}

\newcolumntype{.}{D{.}{.}{-1}}
\newcolumntype{d}[1]{D{.}{.}{#1}}

\colorlet{tableoverheadcolor}{gray!37.5}
\colorlet{tableheadcolor}{gray!25}
\colorlet{tablerowcolor}{gray!12.5}

\newlength{\width}
\newlength{\height}

\def\draftdate{\relax}
\def\mda{\relax}
\def\mua{\relax}
\def\mla{\relax}
\def\draft{
\def\thtystars{******************************}
\def\sixtystars{\thtystars\thtystars}
\typeout{}
\typeout{\sixtystars**}
\typeout{* Draft mode!
         For final version remove \protect\draft\space in source file *}
\typeout{\sixtystars**}
\typeout{}
\def\draftdate{\today}
\def\mua{\marginpar[\boldmath\hfil$\uparrow$]%
                   {\boldmath$\uparrow$\hfil}\color{black}%
                    \typeout{marginpar: $\uparrow$}\ignorespaces}
\def\mda{\color{red}\marginpar[\boldmath\hfil$\downarrow$]%
                   {\boldmath$\downarrow$\hfil}%
                    \typeout{marginpar: $\downarrow$}\ignorespaces}
\def\mla{\marginpar[\boldmath\hfil$\rightarrow$]%
                   {\boldmath$\leftarrow $\hfil}%
                    \typeout{marginpar: $\leftrightarrow$}\ignorespaces}
\def\Mua{\marginpar[\boldmath\hfil$\Uparrow$]%
                   {\boldmath$\Uparrow$\hfil}\color{black}%
                    \typeout{marginpar: $\uparrow$}\ignorespaces}
\def\Mda{\color{red}\marginpar[\boldmath\hfil$\Downarrow$]%
                   {\boldmath$\Downarrow$\hfil}%
                    \typeout{marginpar: $\downarrow$}\ignorespaces}
\def\Mla{\marginpar[\boldmath\hfil\textcolor{red}{$\Rightarrow$}]%
                   {\boldmath\textcolor{red}{$\Leftarrow $}\hfil}%
                    \typeout{marginpar: $\leftrightarrow$}\ignorespaces}
\overfullrule 5pt
\oddsidemargin 15mm
\marginparwidth 29mm
}


%% file: article.bbl
\providecommand{\href}[2]{#2}\begingroup\raggedright\begin{thebibliography}{10}

\bibitem{osti_3914478}
A.~Grassellino, S.~Guha, E.~Isaacs, L.~Jehi, M.~Papermaster, M.~Schmidt,
  T.~Crowder, H.~Lamm, Y.~Lu, T.~Roy, {\em et al.},
  \href{http://dx.doi.org/10.2172/3914478}{{\em SCAC Quantum Committee Report}}
  tech. rep., US Department of Energy (USDOE), Washington, DC (United States).
  Office of Science Advisory Committee, 09, 2026.
\newblock \url{https://www.osti.gov/biblio/3914478}.

\bibitem{Bravo-Prieto:2021ehz}
C.~Bravo-Prieto, J.~Baglio, M.~C\`e, A.~Francis, D.~M. Grabowska, and
  S.~Carrazza, {\em {Style-based quantum generative adversarial networks for
  Monte Carlo events}}.
  \href{http://dx.doi.org/10.22331/q-2022-08-17-777}{Quantum {\bf 6} (2022)
  777}, \href{http://arxiv.org/abs/2110.06933}{{\tt arXiv:2110.06933
  [quant-ph]}}.

\bibitem{Agliardi:2022ghn}
G.~Agliardi, M.~Grossi, M.~Pellen, and E.~Prati, {\em {Quantum integration of
  elementary particle processes}}.
  \href{http://dx.doi.org/10.1016/j.physletb.2022.137228}{Phys. Lett. B {\bf
  832} (2022)  137228}, \href{http://arxiv.org/abs/2201.01547}{{\tt
  arXiv:2201.01547 [hep-ph]}}.

\bibitem{Robbiati:2023mbk}
M.~Robbiati, J.~M. Cruz-Martinez, and S.~Carrazza, {\em {Determining
  probability density functions with adiabatic quantum computing}}.
  \href{http://dx.doi.org/10.1007/s42484-024-00228-2}{Quantum Machine
  Intelligence {\bf 7} (2025) no.~1, 5},
  \href{http://arxiv.org/abs/2303.11346}{{\tt arXiv:2303.11346 [quant-ph]}}.

\bibitem{deLejarza:2023qxk}
J.~J.~M. de~Lejarza, M.~Grossi, L.~Cieri, and G.~Rodrigo,
  \href{http://dx.doi.org/10.1109/QCE57702.2023.00071}{``{Quantum Fourier
  Iterative Amplitude Estimation},''} in {\em {2023 International Conference on
  Quantum Computing and Engineering}}.
\newblock IEEE, 5, 2023.
\newblock \href{http://arxiv.org/abs/2305.01686}{{\tt arXiv:2305.01686
  [quant-ph]}}.

\bibitem{Cruz-Martinez:2023vgs}
J.~M. Cruz-Martinez, M.~Robbiati, and S.~Carrazza, {\em {Multi-variable
  integration with a variational quantum circuit}}.
  \href{http://dx.doi.org/10.1088/2058-9565/ad5866}{Quantum Sci. Technol. {\bf
  9} (2024) no.~3, 035053}, \href{http://arxiv.org/abs/2308.05657}{{\tt
  arXiv:2308.05657 [quant-ph]}}.

\bibitem{deLejarza:2024scm}
J.~J.~M. de~Lejarza, D.~F. Renter\'\i{}a-Estrada, M.~Grossi, and G.~Rodrigo,
  {\em {Quantum integration of decay rates at second order in perturbation
  theory}}. \href{http://dx.doi.org/10.1088/2058-9565/ada9c5}{Quantum Sci.
  Technol. {\bf 10} (2025) no.~2, 025026},
  \href{http://arxiv.org/abs/2409.12236}{{\tt arXiv:2409.12236 [quant-ph]}}.

\bibitem{Williams:2025hza}
I.~Williams and M.~Pellen, {\em {A general approach to quantum integration of
  cross sections in high-energy physics}}.
  \href{http://dx.doi.org/10.1088/2058-9565/adf771}{Quantum Sci. Technol. {\bf
  10} (2025) no.~4, 045017}, \href{http://arxiv.org/abs/2502.14647}{{\tt
  arXiv:2502.14647 [quant-ph]}}.

\bibitem{Lee:2025lzz}
E.~Lee, S.~Lee, and S.~Kim, {\em {Quantum walk based Monte~Carlo simulation for
  photon interaction cross sections}}.
  \href{http://dx.doi.org/10.1103/PhysRevD.111.116001}{Phys. Rev. D {\bf 111}
  (2025) no.~11, 116001}, \href{http://arxiv.org/abs/2502.14374}{{\tt
  arXiv:2502.14374 [quant-ph]}}.

\bibitem{Bauer:2019qxa}
C.~W. Bauer, W.~A. de~Jong, B.~Nachman, and D.~Provasoli, {\em {Quantum
  Algorithm for High Energy Physics Simulations}}.
  \href{http://dx.doi.org/10.1103/PhysRevLett.126.062001}{Phys. Rev. Lett. {\bf
  126} (2021) no.~6, 062001}, \href{http://arxiv.org/abs/1904.03196}{{\tt
  arXiv:1904.03196 [hep-ph]}}.

\bibitem{Bepari:2020xqi}
K.~Bepari, S.~Malik, M.~Spannowsky, and S.~Williams, {\em {Towards a quantum
  computing algorithm for helicity amplitudes and parton showers}}.
  \href{http://dx.doi.org/10.1103/PhysRevD.103.076020}{Phys. Rev. D {\bf 103}
  (2021) no.~7, 076020}, \href{http://arxiv.org/abs/2010.00046}{{\tt
  arXiv:2010.00046 [hep-ph]}}.

\bibitem{Bepari:2021kwv}
K.~Bepari, S.~Malik, M.~Spannowsky, and S.~Williams, {\em {Quantum walk
  approach to simulating parton showers}}.
  \href{http://dx.doi.org/10.1103/PhysRevD.106.056002}{Phys. Rev. D {\bf 106}
  (2022) no.~5, 056002}, \href{http://arxiv.org/abs/2109.13975}{{\tt
  arXiv:2109.13975 [hep-ph]}}.

\bibitem{Deliyannis:2022uyh}
P.~Deliyannis, J.~Sud, D.~Chamaki, Z.~Webb-Mack, C.~W. Bauer, and B.~Nachman,
  {\em {Improving quantum simulation efficiency of final state radiation with
  dynamic quantum circuits}}.
  \href{http://dx.doi.org/10.1103/PhysRevD.106.036007}{Phys. Rev. D {\bf 106}
  (2022) no.~3, 036007}, \href{http://arxiv.org/abs/2203.10018}{{\tt
  arXiv:2203.10018 [hep-ph]}}.

\bibitem{Gustafson:2022dsq}
G.~Gustafson, S.~Prestel, M.~Spannowsky, and S.~Williams, {\em {Collider events
  on a quantum computer}}.
  \href{http://dx.doi.org/10.1007/JHEP11(2022)035}{JHEP {\bf 11} (2022)  035},
  \href{http://arxiv.org/abs/2207.10694}{{\tt arXiv:2207.10694 [hep-ph]}}.

\bibitem{Bauer:2023ujy}
C.~W. Bauer, S.~Chigusa, and M.~Yamazaki, {\em {Quantum parton shower with
  kinematics}}. \href{http://dx.doi.org/10.1103/PhysRevA.109.032432}{Phys. Rev.
  A {\bf 109} (2024) no.~3, 032432},
  \href{http://arxiv.org/abs/2310.19881}{{\tt arXiv:2310.19881 [hep-ph]}}.

\bibitem{Perez-Salinas:2020nem}
A.~P\'erez-Salinas, J.~Cruz-Martinez, A.~A. Alhajri, and S.~Carrazza, {\em
  {Determining the proton content with a quantum computer}}.
  \href{http://dx.doi.org/10.1103/PhysRevD.103.034027}{Phys. Rev. D {\bf 103}
  (2021) no.~3, 034027}, \href{http://arxiv.org/abs/2011.13934}{{\tt
  arXiv:2011.13934 [hep-ph]}}.

\bibitem{Li:2021kcs}
{\bf QuNu} Collaboration, T.~Li, X.~Guo, W.~K. Lai, X.~Liu, E.~Wang, H.~Xing,
  D.-B. Zhang, and S.-L. Zhu, {\em {Partonic collinear structure by quantum
  computing}}. \href{http://dx.doi.org/10.1103/PhysRevD.105.L111502}{Phys. Rev.
  D {\bf 105} (2022) no.~11, L111502},
  \href{http://arxiv.org/abs/2106.03865}{{\tt arXiv:2106.03865 [hep-ph]}}.

\bibitem{deLejarza:2025upd}
J.~J.~M. de~Lejarza, H.-Y. Wu, O.~Kyriienko, G.~Rodrigo, and M.~Grossi, {\em
  {Quantum Chebyshev probabilistic models for fragmentation functions}}.
  \href{http://dx.doi.org/10.1038/s42005-025-02361-1}{Commun. Phys. {\bf 8}
  (2025) no.~1, 448}, \href{http://arxiv.org/abs/2503.16073}{{\tt
  arXiv:2503.16073 [quant-ph]}}.

\bibitem{Chawdhry:2023jks}
H.~A. Chawdhry and M.~Pellen, {\em {Quantum simulation of colour in
  perturbative quantum chromodynamics}}.
  \href{http://dx.doi.org/10.21468/SciPostPhys.15.5.205}{SciPost Phys. {\bf 15}
  (2023) no.~5, 205}, \href{http://arxiv.org/abs/2303.04818}{{\tt
  arXiv:2303.04818 [hep-ph]}}.

\bibitem{Chawdhry:2025iuz}
H.~A. Chawdhry, M.~Pellen, and S.~Williams, {\em {Quantum simulation of
  scattering amplitudes and interferences in perturbative QCD}}.
  \href{http://dx.doi.org/10.1140/epjc/s10052-026-16121-0}{Eur. Phys. J. C {\bf
  86} (2026) no.~8, 961}, \href{http://arxiv.org/abs/2507.07194}{{\tt
  arXiv:2507.07194 [hep-ph]}}.

\bibitem{Bashore:2025uwb}
E.~Bashore, S.~Moretti, and T.~Vitos, {\em {A quantum algorithm for the n-gluon
  MHV scattering amplitude}}.
  \href{http://dx.doi.org/10.21468/SciPostPhys.20.4.114}{SciPost Phys. {\bf 20}
  (2026) no.~4, 114}, \href{http://arxiv.org/abs/2507.14252}{{\tt
  arXiv:2507.14252 [hep-ph]}}.

\bibitem{Haddad:2026ncx}
Y.~Haddad, K.~Xu, V.~Croft, J.~C. Halimeh, and M.~Grossi, {\em {Coherent
  Quantum Evaluation of Collider Amplitudes for Effective Field Theory
  Constraints}}. \href{http://arxiv.org/abs/2602.21311}{{\tt arXiv:2602.21311
  [hep-ph]}}.

\bibitem{Bashore:2026ubq}
E.~Bashore, S.~Moretti, and T.~Vitos, {\em {Quantum computation of partonic
  Drell-Yan scattering cross sections and interference effects}}.
  \href{http://arxiv.org/abs/2608.28057}{{\tt arXiv:2608.28057 [hep-ph]}}.

\bibitem{Ramirez-Uribe:2021ubp}
S.~Ram\'\i{}rez-Uribe, A.~E. Renter\'\i{}a-Olivo, G.~Rodrigo, G.~F.~R.
  Sborlini, and L.~Vale~Silva, {\em {Quantum algorithm for Feynman loop
  integrals}}. \href{http://dx.doi.org/10.1007/JHEP05(2022)100}{JHEP {\bf 05}
  (2022)  100}, \href{http://arxiv.org/abs/2105.08703}{{\tt arXiv:2105.08703
  [hep-ph]}}.

\bibitem{Clemente:2022nll}
G.~Clemente, A.~Crippa, K.~Jansen, S.~Ram\'\i{}rez-Uribe, A.~E.
  Renter\'\i{}a-Olivo, G.~Rodrigo, G.~F.~R. Sborlini, and L.~Vale~Silva, {\em
  {Variational quantum eigensolver for causal loop Feynman diagrams and
  directed acyclic graphs}}.
  \href{http://dx.doi.org/10.1103/PhysRevD.108.096035}{Phys. Rev. D {\bf 108}
  (2023) no.~9, 096035}, \href{http://arxiv.org/abs/2210.13240}{{\tt
  arXiv:2210.13240 [hep-ph]}}.

\bibitem{deLejarza:2024pgk}
J.~J.~M. de~Lejarza, L.~Cieri, M.~Grossi, S.~Vallecorsa, and G.~Rodrigo, {\em
  {Loop Feynman integration on a quantum computer}}.
  \href{http://dx.doi.org/10.1103/PhysRevD.110.074031}{Phys. Rev. D {\bf 110}
  (2024) no.~7, 074031}, \href{http://arxiv.org/abs/2401.03023}{{\tt
  arXiv:2401.03023 [hep-ph]}}.

\bibitem{Ramirez-Uribe:2024wua}
S.~Ram\'\i{}rez-Uribe, A.~E. Renter\'\i{}a-Olivo, and G.~Rodrigo, {\em {Quantum
  querying based on multicontrolled Toffoli gates for causal Feynman loop
  configurations and directed acyclic graphs}}.
  \href{http://arxiv.org/abs/2404.03544}{{\tt arXiv:2404.03544 [quant-ph]}}.

\bibitem{Ochoa-Oregon:2025opz}
S.~A. Ochoa-Oregon, J.~P. Uribe-Ram{\'\i}rez, R.~J. Hern{\'a}ndez-Pinto,
  S.~Ram{\'\i}rez-Uribe, and G.~Rodrigo, {\em {Graph theory-based automated
  quantum algorithm for efficient querying of acyclic and multiloop causal
  configurations}}. \href{http://dx.doi.org/10.1038/s42005-026-02672-x}{Commun.
  Phys. {\bf 9} (2026) no.~1, 270}, \href{http://arxiv.org/abs/2508.04019}{{\tt
  arXiv:2508.04019 [quant-ph]}}.

\bibitem{Draper:2025nku}
P.~Draper, L.~Hidalgo, and A.~Ilderton, {\em {Hamiltonian truncation and
  quantum simulation of strong-field QED beyond tree level}}.
  \href{http://dx.doi.org/10.1103/4lmx-psvz}{Phys. Rev. D {\bf 113} (2026)
  no.~5, 056010}, \href{http://arxiv.org/abs/2509.15495}{{\tt arXiv:2509.15495
  [hep-ph]}}.

\end{thebibliography}\endgroup
